\documentclass[namedreferences]{solarphysics}

\usepackage[hyperref,optionalrh]{spr-sola-addons}
\usepackage{graphicx}        
\usepackage{color}           
\usepackage{breakurl}        

\renewcommand{\vec}[1]{{\mathbfit #1}}

\chardef\us=`\_

\begin{document}

\begin{article}
\begin{opening}

\title{Synchronized Helicity Oscillations: A Link
Between Planetary Tides and the Solar Cycle?}

\author[corref,email={F.Stefani@hzdr.de}]{\inits{F.}\fnm{F.}~\lnm{Stefani}}
\author{\inits{A.}\fnm{A.}~\lnm{Giesecke}}
\author{\inits{N.}\fnm{N.}~\lnm{Weber}}
\author{\inits{T.}\fnm{T.}~\lnm{Weier}}
\address{Helmholtz-Zentrum Dresden -- Rossendorf, P.O. Box 510119,
D-01314 Dresden, Germany}

\runningauthor{F. Stefani {\it et al.}}
\runningtitle{Synchronized Helicity Oscillations: A Possible Link
Between Planetary Orbits and Solar Cycle?}

\begin{abstract}
Recent years have seen an increased interest in the 
question of whether the gravitational action of planets 
could have an influence on the solar dynamo. Without
discussing the observational validity of the claimed
correlations, we ask for a possible physical mechanism 
that might link the weak planetary forces
with solar dynamo action. We focus on the
helicity oscillations that were recently found in simulations
of the current-driven, kink-type Tayler instability, 
which is characterized by an $m=1$ azimuthal dependence.
We show how these helicity oscillations can be resonantly 
excited by some $m=2$ perturbation that reflects a 
tidal oscillation. Specifically, we 
speculate that the 11.07 years tidal oscillation induced by the 
Venus--Earth--Jupiter system may lead to a 1:1 resonant 
excitation of the oscillation of the $\alpha$-effect. 
Finally, in the framework of a reduced, zero-dimensional 
$\alpha$--$\Omega$ dynamo model we recover a 22.14-year
cycle of the solar dynamo.

\end{abstract}
\end{opening}

\section{Introduction}
     \label{S-Introduction}

Sixty years after the seminal article of  \cite{Parker1955},
a remarkable agreement has been achieved that the
solar magnetic field is generated by  some sort of an
$\alpha$--$\Omega$ dynamo \citep{Charbonneau2010}.

Most certainly, the (strong) toroidal field is 
produced from some (weak) poloidal field by
an $\Omega$-effect due to differential rotation, whereas 
the poloidal field is reproduced from the toroidal field by
some appropriate $\alpha$-effect.
The remaining controversy concerns the 
source, and location, of this $\alpha$-effect that
is needed to close the dynamo loop.

Roughly, we can distinguish between four different
interpretations of the toroidal-to-poloidal field transformation. 
Mean-field dynamo theory, focusing on helical twisting of 
toroidal field lines 
in the turbulent convective zone, can be traced back to 
heuristic arguments by \cite{Parker1955} and was later 
corroborated in mathematical detail by \cite{Steenbeck1966}; see 
also \cite{Krause1980}. 
The theory starts by expressing the flow 
${\vec U}=\overline{\vec{U}}+\vec{u}$
and the
magnetic field ${\vec B}=\overline{\vec{B}}+\vec{b}$ 
as the sum of their mean parts (denoted by an overbar) 
and their fluctuating parts  (denoted by lower-case letters).
The interaction of the fluctuating flow and 
magnetic-field components produces 
an additional electromotive force term 
in the induction equation 
which, in its simplest form, can be written as
$\cal{E}=\overline{ \vec{u} \times \vec{b}}=\alpha 
\overline{\vec{B}} -\beta \nabla \times 
\overline{\vec{B}}$ (but see 
\cite{Krause1980} and   \cite{Raedler2006} for significant extensions).

Despite various conceptual problems \citep{Proctor2006}, 
regarding {\it e.g.} 
the catastrophic quenching of $\alpha$
\citep{Vainshtein1992}, or
the questionable relationship between helicity 
and $\alpha$ and the non-convergence of $\alpha$ 
for large magnetic Reynolds 
numbers \citep{Courvoisier2006},
mean-field theory has served for decades as the 
standard model of the solar 
dynamo, which provided
a natural explanation for the periodicity and 
the equator-ward sunspot propagation of the solar cycle
\citep{Steenbeck1969,Stix1972}. 
A first blow to this model came when helioseismology
mapped the differential rotation in the solar interior 
\citep{Brown1989}, in particular the positive radial shear
in a  $\pm 30^{\circ}$ strip around the Equator, 
resulting
in a serious problem with the Parker--Yoshimura sign rule 
which requires $\alpha \partial \Omega/\partial r<0$
in the northern hemisphere for the correct
equator-ward propagation of sun spots 
\citep{Parker1955,Yoshimura1975}. 
A second issue was raised by
\cite{Choudhuri1993} who noticed that the rather 
strong toroidal field
at the bottom of the convection zone, which is needed to 
explain the variation of the tilts of bipolar sunspot pairs with 
latitude, 
would significantly hamper the helical turbulence to twist 
the toroidal field.

A possible way out of this dilemma was found in the 
Babcock--Leighton mechanism \citep{Babcock1961,Leighton1964}, which 
interprets the generation of poloidal field by 
the stronger diffusive cancellation 
of the (closer to the equator) leading sunspots compared with 
that of the trailing (farther from the  equator) spots. 
This leads to 
a spatially separated, or flux-transport 
type of dynamo \citep{Choudhuri1995}, which also provides the correct 
butterfly diagram if combined with an appropriate meridional circulation. 
Most notable in the context of our work is that
the 22-year (Hale) cycle is basically set by the velocity of the
meridional
circulation \citep{CharbonneauDikpati2000}.
The flux-transport dynamo model of 
\cite{Jiang2007} succeeded in predicting the relatively weak 
Solar Cycle 24.
Further to this, a tuned model of this kind
was shown to produce a Maunder-like minimum \citep{Choudhuri2009}, 
although some additional subsurface $\alpha$-effect has still
to be invoked for the dynamo to restart after the minimum 
(when nearly all sunspots had disappeared).

The tachocline $\alpha$-effect proposed by  \cite{Dikpati2001} might
serve this purpose well. It relies on a hydrodynamic shear instability
at the solar tachocline where vertical fluid displacements correlate with 
horizontal-vorticity pattern
providing kinetic helicity, and therefore, a third 
possibility to produce an $\alpha$-effect. 
An alternative version of a distributed dynamo 
implies a stronger role of the near-surface shear layer
that may also be compatible with 
the observed angular velocities 
of magnetic tracers \citep{Brandenburg2005}. 

A fourth interpretation of the toroidal-to-poloidal transformation
relies on the idea that 
the toroidal field itself becomes unstable to non-axisymmetric 
instabilities, which can then lead to an $\alpha$-effect.
With dedicated application to the Sun, this theory has been
worked out by \cite{Ferrizmas1994} and \cite{Zhang2003}.
Such dynamo models, which rely on flux tube instabilities, 
were successfully applied to explain grand minima in terms of on-off 
intermittency \citep{Schmitt1996}.
Interestingly, the underlying current-driven, 
kink-type Tayler instability (TI) 
had been treated long before  
\citep{Tayler1973,Pittstayler1985}.
Recent theoretical \citep{Gellert2011} 
and experimental work \citep{Seilmayer2012} has focused on the TI
in fluids with low magnetic Prandtl number, which indeed 
applies to the solar tachocline.
Based on the TI, a non-linear 
dynamo mechanism had been proposed \citep{Spruit2002} 
which is now known as the ``Tayler--Spruit dynamo''.
However, the initial enthusiasm about  this dynamo 
cooled down with the argument by \cite{Zahn2007}
that the non-axisymmetric $m=1$ TI mode would produce the
``wrong''  poloidal field, 
which is unsuitable for regenerating the dominant 
axisymmetric 
toroidal field.

In principle, this mismatch could be circumvented if the 
$m=1$ TI were connected with some $m=0$ component 
of the $\alpha$-effect.
For comparable large values
of the magnetic Prandtl number [$Pm$], 
\cite{Chatterjee2011}, \cite{Gellert2011}, and 
\cite{Bonanno2012} 
found evidence for spontaneous symmetry breaking between 
left- and right-handed TI modes, leading
indeed to a finite value of $\alpha$.
Whether, and how,  these results  can be transferred 
to the solar tachocline with its
relatively low $Pm\approx 10^{-2}$ will be discussed
further below.

Somewhat disconnected from that main road of solar-dynamo 
research, a few studies were devoted to the theoretical
possibility that the motion of planets could have an 
influence on the solar magnetic field 
\citep{Abreu2012,Charvatova1997,Jose1965,Hung2007,Palus2000,Scafetta2014,Wilson2013}. 
A recent example is the 
article by \cite{Abreu2012} who had 
found synchronized cycles in 
proxies of the solar activity and 
the planetary torques, with periodicities that 
remain phase-locked over 9400 years. Given the immense
relevance of a putative planetary influence on the solar
dynamo and, perhaps, on the Earth's climate and its predictability 
(see Figure 9 in \cite{Scafetta2014}) via
several proposed mechanisms \citep{Svensmark1997,Scafetta2010,Gray2010},
it is not surprising that those claims are vigorously debated.

Yet, first attempts to link solar variability to 
planetary motion
trace back to times of a
milder ``climate'' of scientific disputation.
Noteworthy here is the early article by 
\cite{Bollinger1952},
who showed remarkable evidence of the synchronization of the
sunspot numbers with the 
Venus--Earth--Jupiter system, which is characterized by 
a 44.77-year conjunction cycle. This connection has found further 
attention by \cite{Takahashi1968}, \cite{Wood1972}, 
\cite{CondonSchmidt1975},
\cite{Hung2007}, \cite{Wilson2013},  
and \cite{Okhlopkov2014} who derived the 
actual 
11.07-year  period of the tidal height, 
which is within the 0.1 per cent 
uncertainty of the measured sunspot number period 
of 11.06 years \citep{Cole1973} 
(the slight difference between 11.07 years and
44.77/4=11.19 years 
is a typical aliasing effect
\citep{CondonSchmidt1975}).

Surveying the literature we can distinguish
between studies \citep{Abreu2012,Scafetta2014} 
advocating a
planetary {\it modulation} of the
solar cycle, while accepting the explanatory power of traditional
dynamo models for the 22-year Hale cycle,  
and other studies that, more radically,
relate the Hale cycle to planetary motion, mostly 
to the tidal effect of the Venus--Earth--Jupiter system 
\citep{Bollinger1952,Takahashi1968,Wood1972,Hung2007,Wilson2013,Okhlopkov2014}.

In either case, it is not surprising that the first 
reaction of most scientists is profound skepticism
(if not complete rejection), 
given the tiny accelerations exerted by planets on 
the Sun ($\approx 10^{-10}$ m s$^{-2}$, see  
\cite{DeJager2005} and \cite{Callebaut2012}) and the 
corresponding tidal heights  of less than a mm
\citep{CondonSchmidt1975}.
This being said, one should likewise bear in mind 
the many examples  that show
that very weak forces can indeed lead to synchronization
if only the time of interaction is long enough \citep{Pikovsky}. 

Although the  empirical correlation of the solar cycle with the 
Venus--Earth--Jupiter conjunction cycle seems amazingly persuasive 
(see Figure 1 of  \cite{Bollinger1952}, Figures 1 and 2 of  \cite{Wood1972},
and Figure 3 of \cite{Okhlopkov2014}), 
we will abstain here 
from any judgment of empirical correlations, in particular 
with regard to longer periodicities as discussed 
by \cite{Abreu2012} and in the reactions to 
that article. 

Instead, the aim of this investigation is to explore whether a 
specific mechanism could indeed lead to synchronization  
of the solar dynamo with planetary motion.
The chosen model is mainly inspired by a recent numerical 
finding \citep{Weber2015}
that the TI at  
low magnetic Prandtl numbers is capable of producing  
oscillations of the helicity and the related $\alpha$-effect.
These oscillations are connected with a redistribution  of 
energy between left- and right-handed TI modes, without (or only 
slightly) changing 
the total energy content of the system. We strongly emphasize this 
latter feature
because it indicates the possible point of vantage for the 
tiny planetary forces to synchronize the solar dynamo.

Rather than attacking the complete solar-dynamo problem 
within a single numerical model, we split our argument into two
parts. First, we will show how
helicity oscillations of the $m=1$ TI modes can be resonantly 
excited by some $m=2$ periodic viscosity modulations that serve as a 
surrogate for the corresponding tidal oscillations. Specifically, we 
argue that the 11.07-year tidal oscillation, connected with the
44.77-year conjunction cycle of the 
Venus--Earth--Jupiter system, may lead to a resonant excitation of a
11.07-year oscillation of the $\alpha$-effect. 

Second, in the framework of a strongly reduced, zero-dimensional 
$\alpha$--$\Omega$ dynamo model,  we will recover from this 
11.07-year oscillation of $\alpha$ the 22.14-year
(Hale) cycle, and we will discuss some interesting aspects of this 
model in view of observational features.

The article will conclude with a summary, and a dismayingly
long list of problems that are yet to be solved before 
the proposed  mechanism of planetary synchronization of 
the solar dynamo might get a chance to become accepted.

\section{Resonant Excitation of Helicity Oscillations}

Admittedly, the setting in this section is still
far away from any realistic model of the tachocline,
let alone the whole solar dynamo. 
Its intention is just to 
illustrate the main physical idea of this article: a 
resonant excitation of helicity oscillations, arising with 
some frequency in the saturated state of the 
$m=1$ TI \citep{Weber2015}, by an 
$m=2$ perturbation oscillating with the same frequency.
Actually, similar resonance phenomena have been
discussed in connection with the swing
excitation of galactic dynamos \citep{Chiba1990} 
and with the 
von-K\'{a}rm\'{a}n-sodium (VKS) dynamo experiment \citep{Giesecke2012}.

The typical time-scale we have in mind 
is the 11.07-year tidal oscillation induced by the 
Venus--Earth--Jupiter system, which might
trigger a corresponding oscillation of the helicity and 
the $\alpha$-effect.

For this purpose, we consider a fluid with 
conductivity $\sigma$, viscosity $\nu$, and density 
$\rho$ in a cylindrical volume of geometric 
aspect ratio 
$H/2R=1.25$, threaded by 
an electrical current [$J_0$] of constant density (actually, a more 
tachocline-shaped hollow cylinder, or a thin spherical shell, would 
require much stronger magnetic fields leading to 
significantly higher numerical costs). In the first
instance we also skip any rotation of the fluid
(see the discussion in the conclusions). 
We further choose a much too 
small magnetic Prandtl number $Pm \equiv \mu_0 \sigma \nu=10^{-6}$
(instead of $\approx 10^{-2}$ as for the tachocline),
which is required to ensure the validity of the
applied numerical scheme that is based on the 
quasistatic approximation.
Details of the numerical implementation of the
TI problem can be found in the
appendix, as well as in \cite{Weber2013,Weber2015}.

Without any perturbation, when choosing a sufficiently large 
Hartmann number 
$Ha\equiv \mu_0 J_0 (2\pi)^{-1} (\sigma/(\nu \rho))^{1/2}=100$,
this setting leads to
an intrinsic helicity oscillation with a 
certain frequency $f=1/T_0$ 
that is comparable with the growth rate of the TI 
in its kinematic phase (see Figures 5 and 7 in 
\cite{Weber2015}). Yet, in contrast to this 
well-defined frequency, 
the amplitude of the helicity oscillation is 
very sensitive to the details of the numerical 
implementation, in particular the grid spacing.  

\begin{figure}
\centerline{\includegraphics[width=0.6\textwidth]{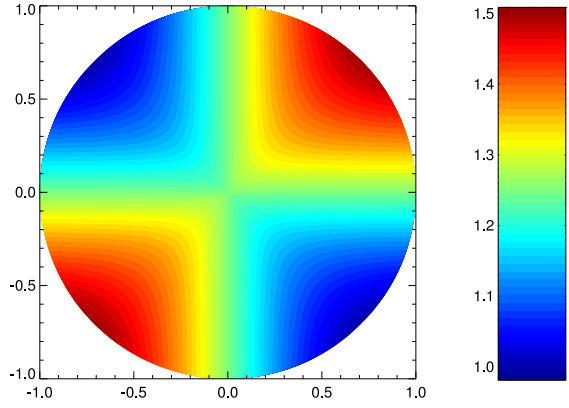}
              }
              \caption{Visualization of the viscosity structure
	      according to Equation (1), at $t=0$ and with $\nu_0=1$, 
	      $A=0.25$.
	      }
\label{nu}
 \end{figure}

Now assume an $m=2$ oscillation  of the viscosity [$\nu$], 
which is to mimic the tidal deformation of the tachocline
due to the Venus--Earth--Jupiter cycle
(a similar way of emulating tidal driving, by using
an $m=2$ body force, was described by \cite{Cebron2014}).
The space-time dependence is assumed to be
\begin{eqnarray}
\nu(r,\phi,t)&=&\nu_0 \{  1    +  A [   1+ 0.5 r^2/R^2 \sin(2 \phi) (1+  \cos(2 \pi t/T_{\nu}) )     ]   \} \; ,
\end{eqnarray}
which includes a constant term $\nu_0 (1+A)$ and an additional 
term with an $m=2$ azimuthal dependence that is oscillating with
a period $T_\nu$. 
The spatial structure of $\nu$ is illustrated in Figure \ref{nu}.

\begin{figure}
\centerline{\includegraphics[width=0.99\textwidth]{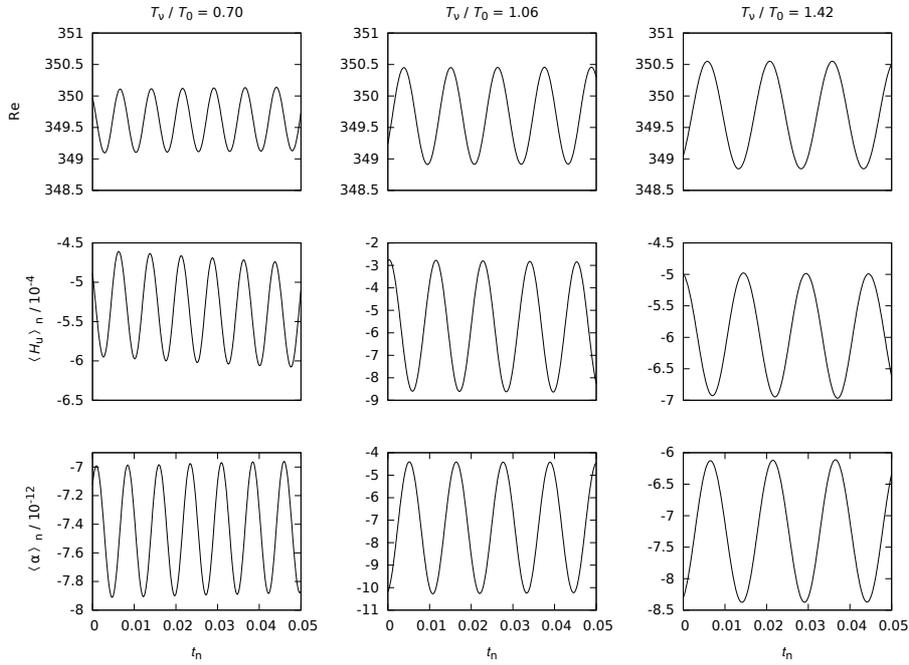}
              }
              \caption{Three examples of the temporal evolution of the 
	      energy, the helicity, and the $\alpha$-effect for  
	      $T_{\nu}/T_0=0.7,1.06,1.42$. The time is normalized 
	      to the viscous 
	      time scale, i.e. $t_n \equiv t \nu/R^2$.
	      }
\label{zeit}
 \end{figure}

\begin{figure}
\centerline{\includegraphics[width=0.8\textwidth]{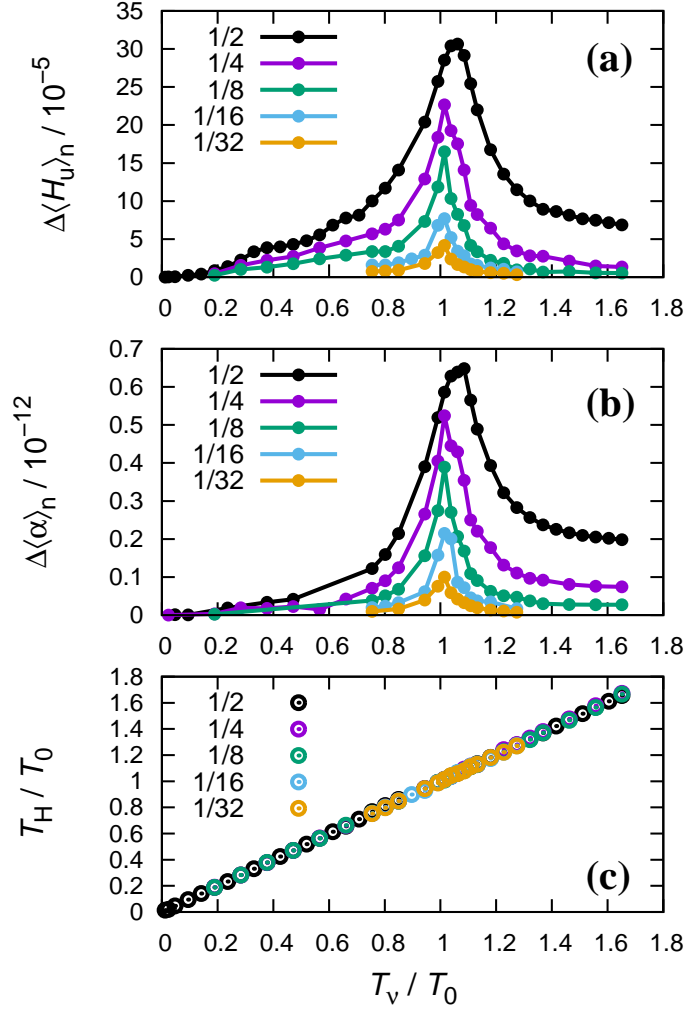}
              }
              \caption{Resonant excitation of the kinetic helicity 
	      and the $\alpha$-effect, 
	      connected with the $m=1$ TI, by the $m=2$ viscosity oscillation
	      with five different amplitudes $A=0.03125, 0.0625, 0.125, 0.25, 0.5$.
	      Amplitude of the oscillations of helicity (a) and  
	      the $\alpha$-effect (b) in dependence
	      on the period of excitation. (c) Oscillation period
	      of helicity in dependence on the period of excitation.
                      }
\label{parametric}
 \end{figure}
 
For the intensity of the viscosity wave we choose now 
five specific values $A=0.03125, 0.0625, 0.125, 0.25, 0.5$.
For $A=0.5$, and three different oscillation periods 
$T_\nu/T_0=0.7, 1.06,$ and 1.42, Figure \ref{zeit} shows the temporal
evolution of three quantities.
The first row gives the  
averaged Reynolds number of the flow arising
from the initial state at rest,
$Re=R (\langle u^2 \rangle)^{1/2}/\nu$ 
where $\langle...\rangle$ 
denotes an average over the total volume. For all three
ratios $T_\nu/T_0$, we get the same $Re \approx 350$ with 
very small fluctuations superposed on it.
In the second row we show the kinetic helicity,
as normalized to the mean-squared velocity over radius, 
i.e.,  ${\langle H_{u} \rangle}_n= \langle {\vec{u}} 
\cdot (\nabla \times {\vec{u}}) \rangle R /\langle u^2 \rangle $.
Apart from some constant part, we observe a significant 
fluctuation (with exactly the period of the viscosity
oscillation), which appears strongest for $T_\nu/T_0=1.06$.
Intimately connected with this helicity oscillation, we
also show (third row) the $\alpha$-effect, normalized in 
such a way
that it corresponds to a magnetic Reynolds number of the 
helical flow part, i.e., 
${\langle \alpha\rangle}_n=\mu_0 \sigma R \langle 
({\vec u} \times {\vec{b}}) \cdot {\vec{B}}_0 \rangle /B^2_0$.

Figure \ref{parametric} shows the amplitude of the 
helicity oscillation (a), the corresponding 
amplitude of the $\alpha$ oscillation (b), and their period (c)
as a function of the period of excitation. 
Evidently we obtain a 
strong resonance (a,b) when the excitation 
frequency $1/T_\nu$ is equal to the intrinsic ``eigenfrequency'' 
$1/T_0$ of the helicity oscillation. 

\begin{figure}
\centerline{\includegraphics[width=0.99\textwidth]{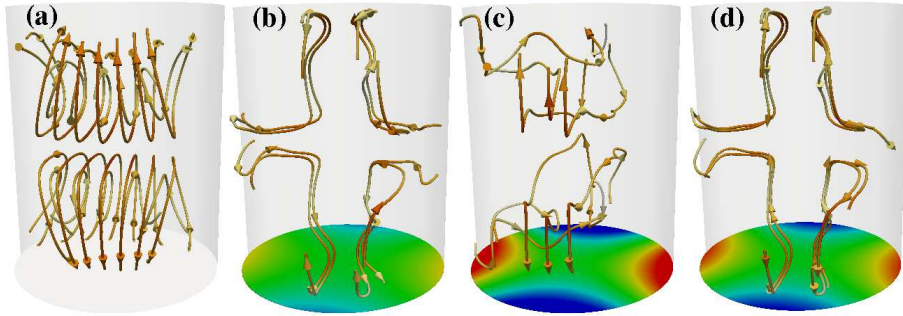}
              }
              \caption{Velocity field over one half-period of the
	      helicity oscillation for the case $A=0.5$, 
	      $T_\nu/T_0=1.06$. (a) Typical $m=1$ velocity field 
	      of the TI,
	      as averaged over one period of the helicity oscillation.
	      (b) Residual velocity (i.e., actual velocity minus
	      averaged velocity), 
	      for an instant with maximum helicity. 
	      (c) Residual velocity for an instant 
	      with mean helicity. (d) Residual velocity 
	      for an instant with minimum helicity.
	      The typical velocities of (a) are a factor of 44 larger 
	      than those of (b,c,d). Note the opposite directions
	      of the residual velocity in (b) and (d). 
	      The color at the bottom
	      of (b,c,d) indicates the viscosity at the respective 
	      instants, according to Equation (1).
                      }
\label{tom}
 \end{figure}

Figure \ref{tom} reveals 
the character of the helicity oscillations
in terms of the velocity for the particular 
case $T_\nu/T_0=1.06$.
While Figure \ref{tom}a illustrates the averaged velocity field,
Figures \ref{tom}b\,--\,d show the residual velocities at 
maximum, mean, and minimum helicity.

\section{Synchronizing the Solar Dynamo}

In the previous section, we have seen that a weak 
$m=2$ viscosity oscillation is able to excite, and 
synchronize, an oscillation of the helicity and the 
corresponding $\alpha$-effect with the same frequency. 

In this section, a first attempt will be made to
set up a closed dynamo model in which the 
synchronized $\alpha$-effect is appropriately embedded.
To keep the physics simple, we will use an extremely reduced, 
zero-dimensional $\alpha$--$\Omega$ dynamo model consisting
of two coupled ordinary differential equations for
the toroidal and the poloidal field component.
Despite their simplicity, models of this sort have been
shown to be well capable of producing various solar-like features 
\citep{Hoyng1993,Weiss2016}, 
in particular if the induction effects in spatially segregated 
layers are mimicked by appropriate time delays in the model
\citep{Wilmotsmith2006}. 

Specifically, we consider the following system of equations:
\begin{eqnarray}   
    \frac{{\rm d} a(t) }{{\rm d} t} &=& \alpha(t) b(t) - a(t)\\
    \frac{{\rm d} b(t)}{{\rm d} t} &=& \Omega a(t) - b(t)
   \end{eqnarray}
wherein $a$ represents the poloidal field (actually, its 
vector potential), and $b$ the toroidal magnetic field. While
keeping constant the value of $\Omega$, which represents the 
induction effect of the differential rotation, $\alpha$ is 
considered as dependent on the instantaneous toroidal
magnetic field:
\begin{eqnarray}   
    \alpha(t) &=& \frac{c}{1+g b^2(t)}
                          +\frac{p b^2(t)}{1+ h b^4(t)} \sin{(2 \pi t/T_{\nu})} \; .
	\label{alpha}		  
   \end{eqnarray}

Equation (\ref{alpha}) is motivated as follows: the first term, 
scaled by $c$,
reflects some constant part that is only quenched, in 
the usual way, by the magnetic-field energy [$b^2$] in 
the denominator. Although this
term can be chosen to be quite small, we will see that it 
must not be set to zero.
The second term, scaled by a parameter 
$p$, is periodic in time and   
emulates the resonance of the $\alpha$-oscillation 
in the sense
that its explicit temporal dependence is fixed, but its 
amplitude has a maximum 
at some particular value of $b$ where the external excitation 
is in resonance with the intrinsic helicity oscillation of 
the TI (note that the frequency of the helicity oscillation is 
a monotonically increasing function of the azimuthal magnetic 
field; see Figure 7 of \cite{Weber2015}). 
Interestingly, a similar $b$-dependence of $\alpha$ had 
already been used by \cite{Wilmotsmith2006}, 
although without 
the $\sin{(2 \pi t/T_{\nu})}$ dependence, and for
other reasons than here.

\begin{figure}
\centerline{\includegraphics[width=0.99\textwidth]{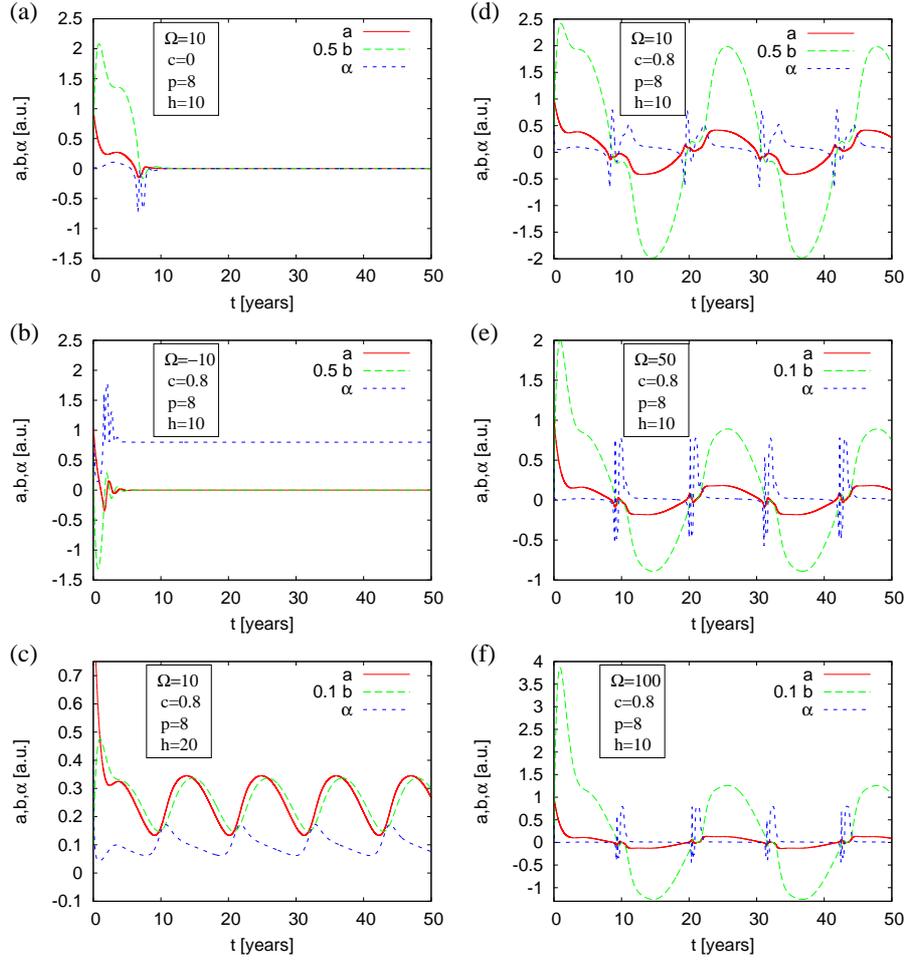}}
              \caption{Evolution of the equation system (2)\,--\,(4) with 
	      $g=1$ and $T_{\nu}=11.07$ years. All other parameters are given in the insets. 
                      }
\label{auto}
 \end{figure}

Figure \ref{auto} illustrates the evolution
of this equation system for six different parameter sets
which cover some paradigmatic types of solutions, although 
not exhaustively. Figures \ref{auto}a\,--\,c correspond to 
solutions that clearly do not comply with 
the solar dynamo, while \ref{auto}d\,--\,f are
much more interesting. In all cases we choose $g=1$
and $T_{\nu}=11.07$ years.

Let us start with Figure \ref{auto}a, obtained for 
$\Omega=10$, $c=0$, $p=8$, and $h=10$. Evidently, this dynamo
fails to work at all, since the constant 
part of $\alpha$ is set exactly to zero.
The dynamo fails also when some constant $\alpha$ is used 
(by setting $c=0.8$), but $\Omega=-10$ acquires the wrong sign;
see Figure \ref{auto}b.
A first dynamo becomes visible in \ref{auto}c, with
$\Omega=10$, $c=0.8$, $p=8$, and a comparably 
large value $h=20$. However, the fields generated by this 
dynamo vacillate with a period of 11.07 years 
around some finite positive values, instead 
of oscillating with the correct 22.14-year period 
around  zero.

Much more promising results are obtained for  
$c=0.8$, $p=8$, $h=10$. For increasing values $\Omega=10, 50, 100$, 
Figures \ref{auto}d\,--\,f show a quite robust solar-type behavior,
with an ever increasing ratio $b/a$ (note the different scaling 
factors for $b$ in the three pictures).
The most important point here is that the 11.07 years (Schwabe)
periodicity of $\alpha$ leads to the 22.14 years  (Hale) 
oscillation of both $a$ and $b$, in contrast to the mere vacillation 
seen in Figure \ref{auto}c.

For the restricted period $27<t<36$ years of Figure \ref{auto}d, 
the next Figure \ref{detail} illustrates in more detail the
behavior during a sign change of the magnetic field, 
including the amazing ``spiky'' 
features of $\alpha$ close to the turning point of $a$ and $b$. 
The capitals A...E mark various instants with specific 
features to be explained in the following:
Initially (A), 
at large values of $b$, $\alpha$ is
rather constant, although strongly quenched, while its
oscillatory part is negligible since $b$ is so strong that 
we are far away from resonance. As $b$ decreases,
it reaches a level at which  
the TI helicity oscillation becomes resonant with the
viscosity oscillation. This happens (B) when $b\approx 0.56$, which 
actually corresponds to 
the maximum of the pre-factor $b^2/(1+10 b^4)$ of the 
oscillatory term in Equation (4).
At this point $\alpha$ becomes strongly negative.
Shortly after (C), $b$ drops to zero,
so that the quenching of the constant term  of $\alpha$  
disappears and $\alpha$ acquires the unquenched value $c$ (here
$\approx 0.8$). Subsequently (D), $b$ goes again through
the resonant point $b\approx -0.56$ for the helicity oscillation 
so that the oscillatory part again contributes its large, 
but now positive, value to $\alpha$. After that (E), $b$ 
increases quite 
smoothly until it reaches a maximum strength where 
$\alpha$ is 
strongly quenched and rather constant.

It is worth noting that many oscillatory
dynamo solutions, based on Equations (2,3) and some appropriate
quenching of  $\alpha$, 
show a similar behavior as long as the quenching of $\alpha$ is
quadratic in the fields. This applies, {\it e.g.}, to 
some of the solutions given by
\cite{Wilmotsmith2006} where a 
close connection with a driven oscillator, in which the driving
works only in a certain window of $b$, has been 
discussed. In some sense, the resonant driving of 
$\alpha$ with the 
11-year cycle can thus be 
considered just a trigger for the whole process.

 \begin{figure}
\centerline{\includegraphics[width=0.8\textwidth]{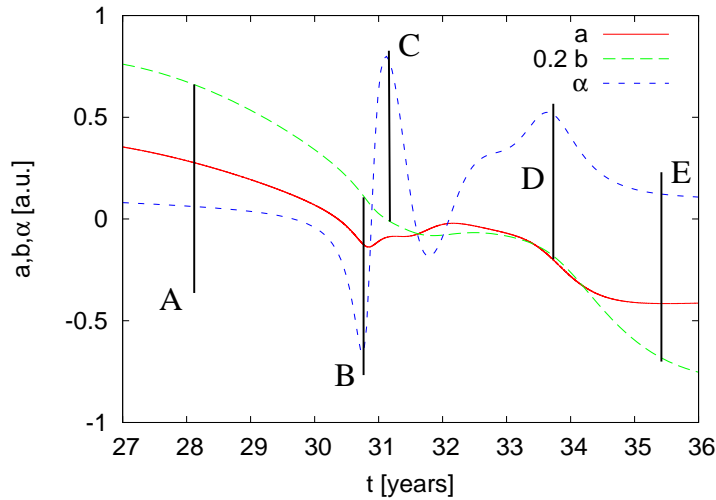}}
              \caption{Details of Figure \ref{auto}d (but note the 
	      different scaling of $b$).
	      For explanations of situations A...E, see the text. 
                      }
\label{detail}
 \end{figure}

\begin{figure}
\centerline{\includegraphics[width=0.8\textwidth]{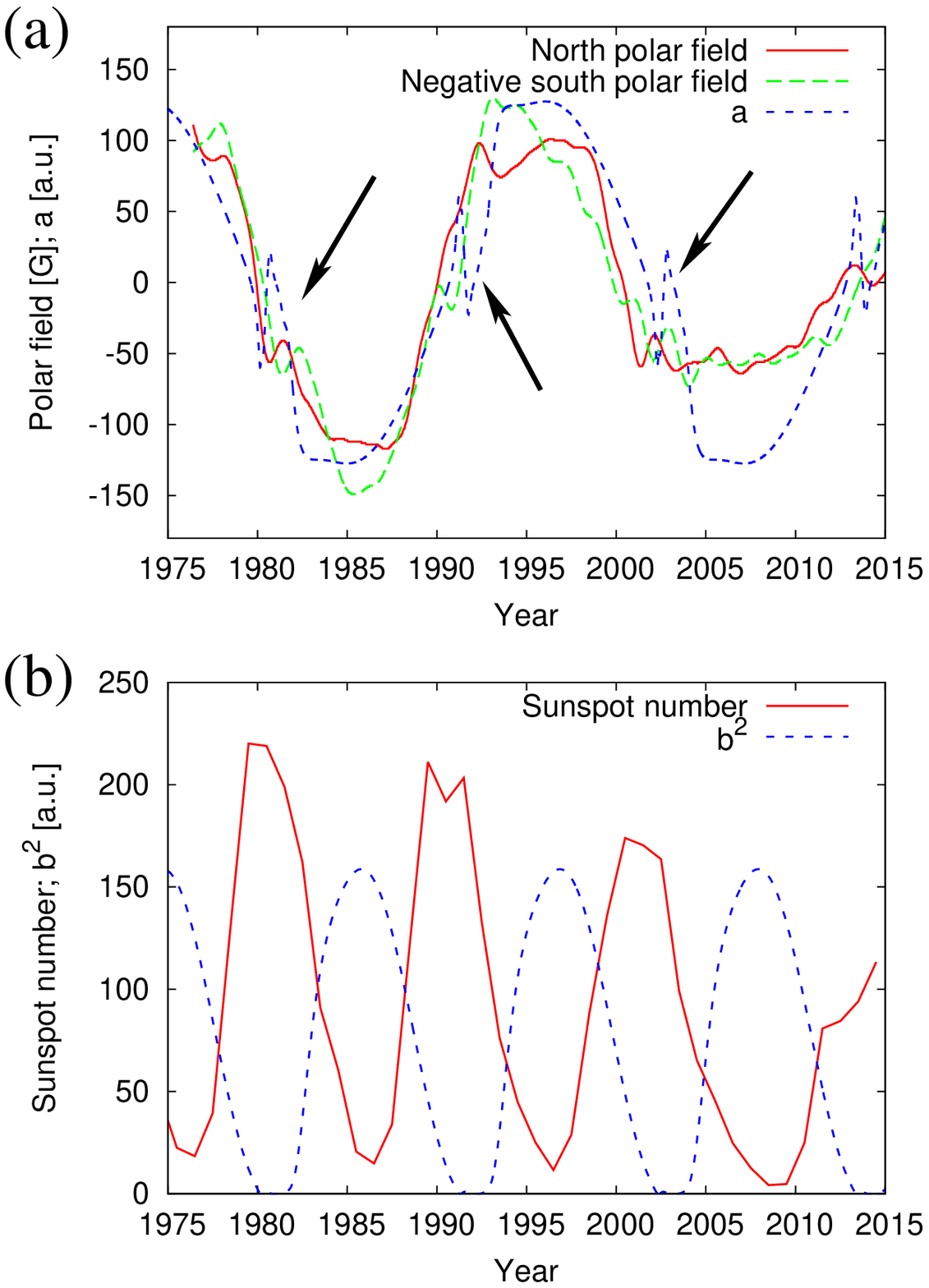}
              }
              \caption{(a) Comparison of the north and south polar magnetic field and the
	      parameter $a$, appropriately scaled and shifted in time. The field data
	      are the 20 nHz filtered data from Wilcox Solar Observatory 
	      (courtesy J.T. Hoeksema). (b) Comparison of the mean sunspot number and
	      the parameter $b^2$, appropriately scaled and shifted in time.
	      The sunspot data are  
	      SILSO data from the Royal Observatory of Belgium, Brussels.
                      }
\label{solar}
 \end{figure}

In Figure \ref{solar} we compare our simulations with two
specific  time
series of the observed solar magnetic field. For this purpose we restrict 
the time  to the period between 1975 and 2015 for which 
north and south polar-field data are available from the
Wilcox Solar Observatory.  Figure \ref{solar}a shows the 20 nHz filtered 
north and south polar-field data, together with an appropriately scaled and
time-shifted segment of our $a(t)$. 
For the same period, Figure \ref{solar}b shows the
annual sunspot number, obtained from the Royal Observatory of Belgium, Brussels,
together with our $b^2(t)$.

The first observation that we make in Figures \ref{auto}, \ref{detail}, and \ref{solar}
is the in-phase behavior 
of $a$ and $b$ which looks not very solar-like at first glance. 
In reality (see Figures \ref{solar}a and b),
the solar magnetic field shows a significant phase-shift between the 
sunspot activity (a tracer for the toroidal field), and the 
global poloidal dipole field.  

A possible way out of this dilemma 
is to invoke the rise-time of the toroidal field, in terms 
of flux tubes, from the 
tachocline to the photosphere. Figure \ref{solar}b  would 
point to a rise-time of approximately four years. 
According to Figure 2  of \cite{Weberfan2013}, this would be
roughly consistent with a magnetic-field strength of 15 kG for
a tube simulation without convection. Unfortunately, the inclusion
of convection leads to rise times not longer than eight months.
Here lies one of the crucial point for the acceptability of this
sort of dynamo localized in, or at least close to the tachocline, while
flux-transport dynamos, or dynamos in which this spatial separation 
is emulated by a time-delay \citep{Wilmotsmith2006} can easily comply with 
this long time-shift.
Only later simulations will show if this discrepancy can be overcome
in higher dimensional $\alpha$--$\Omega$ models.

An amazing coincidence exists, however, between the additional 
peaks of the
north and south polar field in Figure \ref{solar}a and 
the corresponding spikes of our $a$ (indicated by the three black
arrows). 

Another point is related to the vigorous, ''spiky'' 
variations  of $\alpha$ close to the
reversal point of $a$ and $b$, as seen in 
Figure \ref{auto} and Figure \ref{detail}. It is tempting 
to relate this  behavior
to the short-term sign changes of the current-helicity, as 
observed recently by \cite{Zhang2012}. It might also be 
interesting to relate the  high amplitudes of $\alpha$ to the 
so-called $\delta$-sunspots, while other explanations in terms of
kink-type instabilities of flux-tubes have also been invoked to explain 
them \citep{Fan2009}.

While it is not our intention to overemphasize the 
significance of the latter two points, they might be kept 
in mind when trying to validate, or falsify, our
resonant synchronization model of the solar cycle.

\section{Conclusions and Outlook}

While the traditional explanation of the Hale cycle 
bears on intrinsic and, in general, time-independent 
features of the solar dynamo such
as the magnetic diffusivity, the 
amplitudes of $\Omega$, $\alpha$, and the meridional flow 
\citep{CharbonneauDikpati2000},
we have asked for a mechanism that could allow
for synchronizing the solar dynamo with planetary tides. 
From the very outset we were well aware of the fact 
that those tiny forces could never compete with the much 
larger acceleration forces of the turbulent motion in the  
convection zone (if this were indeed 
the governing source of the $\alpha$-effect). The same holds 
true for a corresponding $\alpha$-effect 
based on a Babcock--Leighton mechanism. However,
an additional energy injection from an external forcing is 
{\it not} the crucial point in our argumentation.

Motivated by the recent numerical evidence of helicity oscillations
appearing in the kink-type, $m=1$ 
TI \citep{Weber2015}, we 
studied a simple cylindrical model for the resonant excitation of 
those oscillations by an $m=2$ viscosity variation that 
serves as a proxy for the tidal action of planetary forces. 
Invoking, as a 
specific example, the 11.07-year periodic tide produced by the
Venus--Earth--Jupiter system, this 
was shown to trigger a 11.07-year oscillation
of the helicity and the related $\alpha$-effect. 
This resonant excitation of the $\alpha$-oscillation 
served then as a ``clock'' for the 22.14-year dynamo cycle 
of a reduced, zero-dimensional 
$\alpha-\Omega$ dynamo model. The output of this model shows
interesting solar-like features, in particular an additional
secondary peak of the poloidal field 
shortly after its sign change, and a tendency for 
vigorous and fast oscillations of the helicity in this
weak-field period.

As already mentioned by \cite{Weber2015}, 
there is nearly no external energy needed to 
trigger helicity oscillations. 
As an intrinsic feature of the TI, 
a helicity oscillation is just a reshuffling of 
(kinetic and magnetic) 
energy between left- and right-handed modes, without 
(or barely) changing the energy content, 
as confirmed in Figure \ref{zeit}. 
Exactly here is the point where 
the tiny planetary forces might get a chance to synchronize the 
solar dynamo.
Since this type of dynamo draws its energy nearly exclusively 
from the shear of the differential rotation, the resonantly 
oscillating $\alpha$-effect 
functions as a periodically opening ``bottleneck'' that 
ultimately controls  
the frequency of the dynamo.
An interesting and non-trivial next step would be to check 
if also longer periods of the solar dynamo, like 
the $\approx 87$-year Gleissberg cycle, the $\approx 210$-year 
Suess-de-Vries cycle, 
and the 
$\approx 2300$-year Hallstatt cycle,
can be explained somehow in the framework of the present model.

It goes without saying that the delineated mechanism is by 
far not the end of the story. 
What is urgently needed is to blend together the two  
mechanisms, which were only loosely connected in this article, into 
an appropriate global, i.e. at least a 1D, or better a 2D, 
or 3D dynamo model. 
Perhaps the most significant problem of our model is 
the complete omission of rotation and gravity. The TI 
has been treated only in the presence of 
viscosity and resistivity, so that the typical 
growth rates and frequencies are 
$\propto \omega^2_{\rm{Alfven}}/\omega_{\eta}$ \citep{Ruediger2013}.
This will definitely be modified
when rotation and stratification are taken into account.
Hopefully,  the main result will not change very
dramatically. As shown recently 
\citep{Ruediger2015,Stefani2015},
the effect of positive shear in the tachocline (which prevails 
in a $\pm 30^{\circ}$ strip around the Equator) 
on the TI
is not so grave and leads, surprisingly, even to some 
moderate reduction of the critical Hartmann number.
The azimuthal drift of the instability  mode depends 
strongly on the radial profile of $B_{\phi}(r)$, 
tending to corotate for $B_{\phi}(r)\propto 1/r$
while standing still for $B_{\phi}(r)\propto r$ \citep{Ruediger2015}.

A final remark: Leaving aside the specific TI aspect 
of the helicity oscillations and their
resonant excitation, as discussed in Section 2,
one might ask for other
realizations of the general
resonance model as described in Section 3. Specifically,
it is worthwhile to check whether a similar resonance could 
apply to a tachocline $\alpha$, as proposed by 
\cite{Dikpati2001}. In any case, the  model according to Equations 
(2)\,--\,(4) would then
change significantly due to the missing dependence
of the eigenfrequency on the magnetic-field strength,
which is indeed a specific feature of the TI based 
synchronization model as presented here.

\begin{acks}
This work was supported by the Deutsche Forschungsgemeinschaft 
in the frame of the SPP 1488 (PlanetMag), as well as by 
Helmholtz-Gemeinschaft Deutscher Forschungszentren (HGF)
in frame of the Helmholtz alliance LIMTECH.
Wilcox Solar Observatory data used in this study 
was obtained via the web site 
wso.stanford.edu 
(courtesy of J.T. Hoeksema).
The sunspot data are SILSO data from the Royal Observatory of Belgium, Brussels,
obtained via 
www.sidc.be/silso/infosnytot.
F. Stefani thanks R. Arlt, A. Bonnano, A. Brandenburg, A. Choudhuri, 
D. Hughes, M. Gellert, G. R\"udiger, and 
D. Sokoloff for fruitful discussion
on the solar-dynamo mechanism.
\end{acks}

\section*{Disclosure of Potential Conflicts of Interest}
The authors declare that they have no conflicts of interest. 

\appendix
  \section{The Numerical Model} 

In this appendix we sketch the integro-differential
equation scheme that
was utilized in Section 2 for the calculation of the 
oscillations of the helicity and $\alpha$. Further details can be
found in \cite{Weber2013,Weber2015}. For an alternative 
numerical method to treat the TI, see \cite{Herreman2015}.

In our code we circumvent the usual $Pm$ limitations of 
pure differential-equation codes by replacing the solution of the
induction equation for the 
magnetic field by invoking the so-called quasistatic 
approximation \citep{Davidson2001}. We replace the 
explicit time stepping of 
the magnetic field by computing the electrostatic 
potential by a Poisson solver,
and deriving the electric-current density.
In contrast to many other inductionless approximations
in which this procedure is sufficient,
in our case we cannot avoid to compute the induced 
magnetic field, too. The reason for this is the 
presence of an externally applied electrical current
in the fluid. Computing the Lorentz-force term it turns out that
the product of the applied current with the induced field 
is of the same
order as the product of the magnetic field 
(due to the applied current) with the induced current.
The induced magnetic field is computed as follows: 
in the interior of the domain, we apply the
quasi-stationary approximation and
solve the vectorial Poisson equation for the
magnetic field 
which results when the temporal derivative 
in the induction equation is set to zero.
At the boundary of the domain, however,
the induced magnetic field is  computed
from the induced current density
by means of Biot--Savart's law. This way we arrive at an 
integro-differential
equation approach, similar to the method used by 
\cite{Meir2004}. 

In detail, the numerical model as developed by 
\cite{Weber2013} works as follows: it uses the OpenFOAM library 
to solve the Navier--Stokes 
equations (NSE) for incompressible fluids
\begin{eqnarray}\label{eqn:navierstokes}
\dot {\vec u} + \left({\vec u}\cdot\nabla\right){\vec u} = 
- \nabla p + \nu \Delta {\vec u} + \frac{\vec f_{\mathrm
    L} }{\rho}\hspace{5mm}\textrm{and}\hspace{5mm} \nabla\cdot \vec u = 0,
\end{eqnarray}
with $\vec u$ denoting the velocity, $p$ the (modified) pressure, 
$\vec f_{\mathrm L} = \vec J \times \vec B $ the
electromagnetic Lorentz force density, $\vec J$ the 
total current density, and $\vec B$
the total magnetic field. The NSE
is solved using the PISO algorithm and applying no slip boundary
conditions at the walls. 

Ohm's law in moving conductors
\begin{eqnarray}
{\vec j} = \sigma\left(-\nabla\varphi + {\vec u}\times {\vec B}\right)
\end{eqnarray}
allows us to compute the induced current $\vec j$ by previously
solving a Poisson equation for the perturbed electric potential
$\varphi = \phi -J_0z/\sigma$:
\begin{eqnarray} \Delta\varphi = 
\nabla\cdot\left({\vec u} \times {\vec B}\right).
\end{eqnarray}
We concentrate now on cylindrical 
geometries with an axially
applied current. After 
subtracting the (constant) potential part 
[$J_0z/\sigma$], with $z$ as the
coordinate along the cylinder axis, we use the simple
boundary condition $\varphi = 0$ on top and bottom and $\vec n\cdot \nabla
\varphi=0$ on the mantle of the cylinder, with $\bf n$ as the surface
normal vector. 

The induced magnetic field at the boundary of the domain 
can then be
calculated by Biot--Savart's law
\begin{eqnarray}\label{eqn:biotsavart}
{\vec b}({\vec r}) = \frac{\mu_0}{4\pi}\int dV' \, 
\frac{{\vec j}({\vec r}') \times ({\vec r}-{\vec r}')}{\left|{\vec r}-{\vec r}'\right|^3}.
\end{eqnarray}
In the bulk of the domain, the magnetic field is 
computed by solving
the vectorial Poisson equation
\begin{eqnarray}\label{eqn:induction} 
\Delta {\vec b}=\mu_0 \sigma \nabla \times ( {\vec{u}} \times {\vec{B}} )
\end{eqnarray}
which results from the full time-dependent induction equation
in the quasi-stationary approximation.

Knowing $\vec b$ and $\vec j$ we compute the
Lorentz force ${\vec f}_{\mathrm L}$ for the next iteration.  For more 
details about the numerical scheme, see Section 
2 and 3 of \cite{Weber2013}.

     

\begin{thebibliography}{59}
\ifx\bisbn     \undefined \def\bisbn  #1{ISBN #1}\fi
\ifx\binits    \undefined \def\binits#1{#1}\fi
\ifx\bauthor   \undefined \def\bauthor#1{#1}\fi
\ifx\batitle   \undefined \def\batitle#1{#1}\fi
\ifx\bjtitle   \undefined \def\bjtitle#1{\textit{#1}}\fi
\ifx\bvolume   \undefined \def\bvolume#1{\textbf{#1}}\fi
\ifx\byear     \undefined \def\byear#1{#1}\fi
\ifx\bissue    \undefined \def\bissue#1{#1}\fi
\ifx\bfpage    \undefined \def\bfpage#1{#1}\fi
\ifx\blpage    \undefined \def\blpage #1{#1}\fi
\ifx\burl      \undefined \def\burl#1{\textsf{#1}}\fi
\ifx\href      \undefined \def\href#1#2{\textsf{#2}}\fi
\ifx\betal     \undefined \def\betal{\textit{et al.}}\fi
\ifx\bctitle   \undefined \def\bctitle#1{#1}\fi
\ifx\beditor   \undefined \def\beditor#1{#1}\fi
\ifx\bbtitle   \undefined \def\bbtitle#1{\textit{#1}}\fi
\ifx\bedition  \undefined \def\bedition#1{#1}\fi
\ifx\bseriesno \undefined \def\bseriesno#1{\textbf{#1}}\fi
\ifx\blocation \undefined \def\blocation#1{#1}\fi
\ifx\bsertitle \undefined \def\bsertitle#1{\textit{#1}}\fi
\ifx\bsnm      \undefined \def\bsnm#1{#1}\fi
\ifx\bsuffix   \undefined \def\bsuffix#1{#1}\fi
\ifx\bparticle \undefined \def\bparticle#1{#1}\fi
\ifx\barticle  \undefined \def\barticle#1{}\fi
\ifx\binstitute  \undefined \def\binstitute#1{#1}\fi
\ifx\bpublisher  \undefined \def\bpublisher#1{#1}\fi
\ifx\doiurl    \undefined
  \def\doiurl#1{\href{http://dx.doi.org/#1}{\textsf{DOI}}}\fi
\ifx\arxivurl  \undefined
  \def\arxivurl#1{\href{http://arxiv.org/abs/#1}{\textsf{arXiv}}}\fi
\ifx\adsurl    \undefined
  \def\adsurl#1{\href{http://adsabs.harvard.edu/abs/#1}{\textsf{ADS}}}\fi
\ifx\botherref \undefined \def\botherref#1{}\fi
\ifx\url       \undefined \def\url#1{\textsf{#1}}\fi
\ifx\bchapter  \undefined \def\bchapter#1{}\fi
\ifx\bbook     \undefined \def\bbook#1{}\fi
\ifx\bcomment  \undefined \def\bcomment#1{#1}\fi
\ifx\oauthor   \undefined \def\oauthor#1{#1}\fi
\ifx\citeauthoryear \undefined\def \citeauthoryear#1{#1}\fi
\ifx\endbibitem\undefined \def\endbibitem{}\fi
\ifx\bconflocation  \undefined \def\bconflocation#1{#1} \fi

\bibitem[\protect\citeauthoryear{Abreu \textit{et~al.}}{{2012}}]{Abreu2012}
\begin{barticle}
\bauthor{\bsnm{Abreu}, \binits{J.A.}},
\bauthor{\bsnm{Beer}, \binits{J.}},
\bauthor{\bsnm{Ferriz-Mas}, \binits{A.}},
\bauthor{\bsnm{McCracken}, \binits{K.G.}},
\bauthor{\bsnm{Steinhilber}, \binits{F.}}:
\byear{{2012}},
\batitle{{Is there a planetary influence on solar activity?}}
\bjtitle{{Astron. Astrophys.}}
\bvolume{{548}},
\bfpage{{A88}}
\doiurl{10.1051/0004-6361/201219997}.
\end{barticle}
\endbibitem

\bibitem[\protect\citeauthoryear{Babcock}{1961}]{Babcock1961}
\begin{barticle}
\bauthor{\bsnm{Babcock}, \binits{H.W.}}:
\byear{1961},
\batitle{The topology of the suns magnetic field and the 22-year cycle}.
\bjtitle{Astrophys. J.}
\bvolume{133},
\bfpage{572}
\doiurl{10.1086/147060}.
\end{barticle}
\endbibitem



\bibitem[\protect\citeauthoryear{Bollinger}{1952}]{Bollinger1952}
\begin{barticle}
\bauthor{\bsnm{Bollinger}, \binits{C.J.}}:
\byear{1952},
\batitle{A 44.77 year Jupiter-Venus-Earth configuration sun-tide period in
  solar-climatic cycles}.
\bjtitle{Proc. Okla. Acad. Sci.}
\bvolume{33},
\bfpage{307}.
\end{barticle}
\endbibitem

\bibitem[\protect\citeauthoryear{Bonanno \textit{et~al.}}{2012}]{Bonanno2012}
\begin{barticle}
\bauthor{\bsnm{Bonanno}, \binits{A.}},
\bauthor{\bsnm{Brandenburg}, \binits{A.}},
\bauthor{\bsnm{Del~Sordo}, \binits{F.}},
\bauthor{\bsnm{Mitra}, \binits{D.}}:
\byear{2012},
\batitle{Breakdown of chiral symmetry during saturation of the Tayler
  instability}.
\bjtitle{Phys. Rev. E}
\bvolume{86},
\bfpage{016313}
\doiurl{10.1103/PhysRevE.86.016313}.
\end{barticle}
\endbibitem


\bibitem[\protect\citeauthoryear{Brandenburg}{2005}]{Brandenburg2005}
\begin{barticle}
\bauthor{\bsnm{Brandenburg}, \binits{A.}}:
\byear{2005},
\batitle{The case for a distributed solar dynamo shaped by near-surface shear}.
\bjtitle{Astrophys. J.}
\bvolume{625},
\bfpage{625}
\doiurl{10.1086/429584}.
\end{barticle}
\endbibitem


\bibitem[\protect\citeauthoryear{Brown \textit{et~al.}}{1989}]{Brown1989}
\begin{barticle}
\bauthor{\bsnm{Brown}, \binits{T.M}},
\bauthor{\bsnm{Christensen-Dalsgaard}, \binits{J.}},
\bauthor{\bsnm{Dziembowski}, \binits{W.A.}},
\bauthor{\bsnm{Goode}, \binits{P.}},
\bauthor{\bsnm{Gough}, \binits{D.O.}},
\bauthor{\bsnm{Morrow}, \binits{C.}}:
\byear{1989},
\batitle{Inferring the sun's internal angular velocity from observed 
p-mode frequency splitting}.
\bjtitle{Astrophys. J.}
\bvolume{343},
\bfpage{526}
\doiurl{10.1086/167727}.
\end{barticle}
\endbibitem



\bibitem[\protect\citeauthoryear{Callebaut, de~Jager, and
  Duhau}{2012}]{Callebaut2012}
\begin{barticle}
\bauthor{\bsnm{Callebaut}, \binits{D.K.}},
\bauthor{\bparticle{de} \bsnm{Jager}, \binits{C.}},
\bauthor{\bsnm{Duhau}, \binits{S.}}:
\byear{2012},
\batitle{The influence of planetary attractions on the solar tachocline}.
\bjtitle{J. Atmos. Sol.-Terr. Phys.}
\bvolume{80},
\bfpage{73}
\doiurl{10.1016/j.jastp.2012.03.005}.
\end{barticle}
\endbibitem

\bibitem[\protect\citeauthoryear{Cebron and Hollerbach}{2014}]{Cebron2014}
\begin{botherref}
\oauthor{\bsnm{Cebron}, \binits{D.}},
\oauthor{\bsnm{Hollerbach}, \binits{R.}}:
2014,
{Tidally driven dynamos in a rotating sphere}.
\textit{Astrophys. J. Lett.}
\bvolume{789},
\bfpage{L25}
\doiurl{10.1088/2041-8205/789/1/L25}.
\end{botherref}
\endbibitem

\bibitem[\protect\citeauthoryear{Charbonneau}{{2010}}]{Charbonneau2010}
\begin{botherref}
\oauthor{\bsnm{Charbonneau}, \binits{P.}}:
{2010},
{Dynamo models of the solar cycle}.
\textit{{Liv. Rev. Solar Phys.}}
\bvolume{7},
\bfpage{3}
\doiurl{10.12942/lrsp-2010-3}.
\end{botherref}
\endbibitem

\bibitem[\protect\citeauthoryear{Charbonneau and
  Dikpati}{2000}]{CharbonneauDikpati2000}
\begin{barticle}
\bauthor{\bsnm{Charbonneau}, \binits{P.}},
\bauthor{\bsnm{Dikpati}, \binits{M.}}:
\byear{2000},
\batitle{Stochastic fluctuations in a Babcock-model of the solar cycle}.
\bjtitle{Astrophys. J.}
\bvolume{543},
\bfpage{1027}
\doiurl{10.1086/317142}.
\end{barticle}
\endbibitem

\bibitem[\protect\citeauthoryear{Charvatova}{1997}]{Charvatova1997}
\begin{barticle}
\bauthor{\bsnm{Charvatova}, \binits{I.}}:
\byear{1997},
\batitle{Solar-terrestrial and climatic phenomena in relation to solar inertial
  motion}.
\bjtitle{Surv. Geophys.}
\bvolume{18},
\bfpage{131}
\doiurl{10.1023/A:1006527724221}.
\end{barticle}
\endbibitem

\bibitem[\protect\citeauthoryear{Chatterjee
  \textit{et~al.}}{2011}]{Chatterjee2011}
\begin{barticle}
\bauthor{\bsnm{Chatterjee}, \binits{P.}},
\bauthor{\bsnm{Mitra}, \binits{D.}},
\bauthor{\bsnm{Brandenburg}, \binits{A.}},
\bauthor{\bsnm{Rheinhardt}, \binits{M.}}:
\byear{2011},
\batitle{Spontaneous chiral symmetry breaking by hydromagnetic buoyancy}.
\bjtitle{Phys. Rev. E}
\bvolume{84},
\bfpage{025403}
\doiurl{10.1103/PhysRevE.84.025403}.
\end{barticle}
\endbibitem

\bibitem[\protect\citeauthoryear{Chiba and Tosa}{1990}]{Chiba1990}
\begin{barticle}
\bauthor{\bsnm{Chiba}, \binits{M.}},
\bauthor{\bsnm{Tosa}, \binits{M.}}:
\byear{1990},
\batitle{Swing excitation of galactic magnetic-fields induced by spiral density
  waves}.
\bjtitle{Mon. Not. Roy. Astron. Soc.}
\bvolume{244},
\bfpage{714}
\doiurl{10.1093/mnras/238.2.621}.
\end{barticle}
\endbibitem


\bibitem[\protect\citeauthoryear{Choudhuri, Sch\"ussler, and
  Dikpati}{1995}]{Choudhuri1995}
\begin{barticle}
\bauthor{\bsnm{Choudhuri}, \binits{A.R.}},
\bauthor{\bsnm{Sch\"ussler}, \binits{M.}},
\bauthor{\bsnm{Dikpati}, \binits{M.}}:
\byear{1995},
\batitle{The solar dynamo with meridional circulation}.
\bjtitle{Astron. Astrophys.}
\bvolume{303},
\bfpage{L29}.
\end{barticle}
\endbibitem

\bibitem[\protect\citeauthoryear{Choudhuri and Karak}{2009}]{Choudhuri2009}
\begin{barticle}
\bauthor{\bsnm{Choudhuri}, \binits{A.R.}},
\bauthor{\bsnm{Karak}, \binits{B.B.}}:
\byear{2009},
\batitle{A possible explanation of the Maunder minimum from a flux transport
  dynamo model}.
\bjtitle{Res. Astron. Astrophys.}
\bvolume{9},
\bfpage{953}.
\end{barticle}
\endbibitem

\bibitem[\protect\citeauthoryear{Cole}{1973}]{Cole1973}
\begin{barticle}
\bauthor{\bsnm{Cole}, \binits{T.W.}}:
\byear{1973},
\batitle{Periodicities in solar activities}.
\bjtitle{Solar Phys.}
\bvolume{30},
\bfpage{103}
\doiurl{10.1007/BF00156178}.
\end{barticle}
\endbibitem


\bibitem[\protect\citeauthoryear{Condon and Schmidt}{1975}]{CondonSchmidt1975}
\begin{barticle}
\bauthor{\bsnm{Condon}, \binits{J.J.}},
\bauthor{\bsnm{Schmidt}, \binits{R.R.}}:
\byear{1975},
\batitle{Planetary tides and the sunspot cycles}.
\bjtitle{Solar Phys.}
\bvolume{42},
\bfpage{529}
\doiurl{10.1007/BF00149930}.
\end{barticle}
\endbibitem


\bibitem[\protect\citeauthoryear{Courvoisier, Hughes, and
  Tobias}{2012}]{Courvoisier2006}
\begin{barticle}
\bauthor{\bsnm{Courvoisier}, \binits{A.}},
\bauthor{\bsnm{Hughes}, \binits{D.W.}},
\bauthor{\bsnm{Tobias}, \binits{S.M.}}:
\byear{2006},
\batitle{$\alpha$ effect in a family of chaotic flows}.
\bjtitle{Phys. Rev. Lett.}
\bvolume{96},
\bfpage{034503}
\doiurl{10.1103/PhysRevLett.96.034503}.
\end{barticle}
\endbibitem



\bibitem[\protect\citeauthoryear{Davidson}{2001}]{Davidson2001}
\begin{bbook}
\bauthor{\bsnm{Davidson}, \binits{P.A.}}:
\byear{2001},
\bbtitle{An introduction to magnetohydrodynamics},
\bpublisher{Cambridge University Press}, \blocation{Cambridge}.
\end{bbook}
\endbibitem

\bibitem[\protect\citeauthoryear{De~Jager and Versteegh}{2005}]{DeJager2005}
\begin{barticle}
\bauthor{\bsnm{De~Jager}, \binits{C.}},
\bauthor{\bsnm{Versteegh}, \binits{G.}}:
\byear{2005},
\batitle{Do planetary motions drive solar variability?}
\bjtitle{Solar Phys.}
\bvolume{229},
\bfpage{175}
\doiurl{10.1007/s11207-005-4086-7}.
\end{barticle}
\endbibitem





\bibitem[\protect\citeauthoryear{Dikpati and Gilman}{2001}]{Dikpati2001}
\begin{barticle}
\bauthor{\bsnm{Dikpati}, \binits{M.}},
\bauthor{\bsnm{Gilman}, \binits{P.}}:
\byear{2001},
\batitle{Flux-transport dynamos with alpha-effect from global instability of
  tachocline differential rotation: A solution for magnetic parity selection in
  the sun}.
\bjtitle{Astrophys. J.}
\bvolume{559},
\bfpage{428}
\doiurl{10.1086/322410}.
\end{barticle}
\endbibitem


\bibitem[\protect\citeauthoryear{D'Silva and Choudhuri}{1993}]{Choudhuri1993}
\begin{barticle}
\bauthor{\bsnm{D'Silva}, \binits{S.}},
\bauthor{\bsnm{Choudhuri}, \binits{A.R.}}:
\byear{1993},
\batitle{A theoretical model for tilts of bipolar magnetic regions}.
\bjtitle{Astron. Astrophys.}
\bvolume{272},
\bfpage{621}.
\end{barticle}
\endbibitem

\bibitem[\protect\citeauthoryear{Fan}{{2009}}]{Fan2009}
\begin{botherref}
\oauthor{\bsnm{Fan}, \binits{Y.}}:
{2009},
{Magnetic fields in the solar convection zone}.
\textit{{Liv. Rev. Solar Phys.}}
\bvolume{6},
\bfpage{4}
\doiurl{10.12942/lrsp-2009-4}.
\end{botherref}
\endbibitem

\bibitem[\protect\citeauthoryear{Ferriz~Mas, Schmitt, and
  Sch\"ussler}{1994}]{Ferrizmas1994}
\begin{barticle}
\bauthor{\bsnm{Ferriz~Mas}, \binits{A.}},
\bauthor{\bsnm{Schmitt}, \binits{D.}},
\bauthor{\bsnm{Sch\"ussler}, \binits{M.}}:
\byear{1994},
\batitle{A dynamo effect due to instability of magnetic flux tubes}.
\bjtitle{Astron. Astrophys.}
\bvolume{289},
\bfpage{949}.
\end{barticle}
\endbibitem

\bibitem[\protect\citeauthoryear{Gellert, R\"{u}diger, and
  Hollerbach}{2011}]{Gellert2011}
\begin{barticle}
\bauthor{\bsnm{Gellert}, \binits{M.}},
\bauthor{\bsnm{R\"{u}diger}, \binits{G.}},
\bauthor{\bsnm{Hollerbach}, \binits{R.}}:
\byear{2011},
\batitle{Helicity and alpha-effect by current-driven instabilities of helical
  magnetic fields}.
\bjtitle{Mon. Not. Roy. Astron. Soc.}
\bvolume{414},
\bfpage{2696}
\doiurl{10.1111/j.1365-2966.2011.18583.x}.
\end{barticle}
\endbibitem




\bibitem[\protect\citeauthoryear{Giesecke, Stefani, and
  Burguete}{2012}]{Giesecke2012}
\begin{barticle}
\bauthor{\bsnm{Giesecke}, \binits{A.}},
\bauthor{\bsnm{Stefani}, \binits{F.}},
\bauthor{\bsnm{Burguete}, \binits{J.}}:
\byear{2012},
\batitle{Impact of time-dependent nonaxisymmetric velocity perturbations on
  dynamo action of von K\'{a}rm\'{a}n-like flows}.
\bjtitle{Phys. Rev. E}
\bvolume{86},
\bfpage{066303}
\doiurl{10.1103/PhysRevE.86.066303}.
\end{barticle}
\endbibitem


\bibitem[\protect\citeauthoryear{Gray \textit{et~al.}}{2010}]{Gray2010}
\begin{barticle}
\bauthor{\bsnm{Gray}, \binits{L.J.}},
\bauthor{\bsnm{Beer}, \binits{J.}},
\bauthor{\bsnm{Geller}, \binits{M.}},
\bauthor{\bsnm{Haigh}, \binits{J.D.}},
\bauthor{\bsnm{Lockwood}, \binits{M.}},
\bauthor{\bsnm{Matthes}, \binits{K.}},
\bauthor{\bsnm{Cubasch}, \binits{U.}},
\bauthor{\bsnm{Fleitmann}, \binits{D.}},
\bauthor{\bsnm{Harrison}, \binits{G.}},
\bauthor{\bsnm{Hood}, \binits{L.}},
\bauthor{\bsnm{Luterbacher}, \binits{J.}},
\bauthor{\bsnm{Meehl}, \binits{G.A.}},
\bauthor{\bsnm{Shindell}, \binits{D.}},
\bauthor{\bsnm{van Geel}, \binits{B.}},
\bauthor{\bsnm{White}, \binits{W.}}:
\byear{2010},
\batitle{Solar inluences on climate}.
\bjtitle{Rev. Geophys.}
\bvolume{48},
\bfpage{RG4001}
\doiurl{10.1029/2009RG000282}.
\end{barticle}
\endbibitem


\bibitem[\protect\citeauthoryear{Herreman \textit{et~al.}}{2015}]{Herreman2015}
\begin{barticle}
\bauthor{\bsnm{Herreman}, \binits{W.}},
\bauthor{\bsnm{Nore}, \binits{C.}},
\bauthor{\bsnm{Cappanera}, \binits{L.}},
\bauthor{\bsnm{Guermond}, \binits{J.-L.}}:
\byear{2015},
\batitle{Tayler instability in liquid metal columns and liquid metal
  batteries}.
\bjtitle{J. Fluid Mech.}
\bvolume{771},
\bfpage{79}
\doiurl{10.1017/jfm.2015.159}.
\end{barticle}
\endbibitem

\bibitem[\protect\citeauthoryear{Hoyng}{1993}]{Hoyng1993}
\begin{barticle}
\bauthor{\bsnm{Hoyng}, \binits{P.}}:
\byear{1993},
\batitle{Helicity fluctuations in mean-field theory: an explanation 
for the variability of the solar cycle?}.
\bjtitle{Astron. Astrophys.}
\bvolume{272},
\bfpage{321}.
\end{barticle}
\endbibitem



\bibitem[\protect\citeauthoryear{Hung}{2007}]{Hung2007}
\begin{barticle}
\bauthor{\bsnm{Hung}, \binits{C.-C.}}:
\byear{2007},
\batitle{Apparent relations between solar activity and solar tides caused by
  the planets. NASA/TM}.
\bvolume{2007-214817},
\bfpage{1}.
\end{barticle}
\endbibitem

\bibitem[\protect\citeauthoryear{Jiang, Chatterjee, and
  Choudhuri}{2007}]{Jiang2007}
\begin{barticle}
\bauthor{\bsnm{Jiang}, \binits{J.}},
\bauthor{\bsnm{Chatterjee}, \binits{P.}},
\bauthor{\bsnm{Choudhuri}, \binits{A.R.}}:
\byear{2007},
\batitle{Solar activity forecast with a dynamo model}.
\bjtitle{Mon.  Roy. Astron. Soc.}
\bvolume{381},
\bfpage{1527}
\doiurl{10.1111/j.1365-2966.2007.12267.x}.
\end{barticle}
\endbibitem

\bibitem[\protect\citeauthoryear{Jose}{1965}]{Jose1965}
\begin{barticle}
\bauthor{\bsnm{Jose}, \binits{P.D.}}:
\byear{1965},
\batitle{Suns motion and sunspots}.
\bjtitle{Astron. J.}
\bvolume{70},
\bfpage{193}
\doiurl{10.1086/109714}.
\end{barticle}
\endbibitem




\bibitem[\protect\citeauthoryear{Krause and R{\"a}dler}{1980}]{Krause1980}
\begin{bbook}
\bauthor{\bsnm{Krause}, \binits{F.}},
\bauthor{\bsnm{R{\"a}dler}, \binits{K.-H.}}:
\byear{1980},
\bbtitle{Mean-field magnetohydrodynamics and dynamo theory},
\bpublisher{Akademie-Verlag}, \blocation{Berlin}.
\end{bbook}
\endbibitem

\bibitem[\protect\citeauthoryear{Leighton}{1964}]{Leighton1964}
\begin{barticle}
\bauthor{\bsnm{Leighton}, \binits{R.B.}}:
\byear{1964},
\batitle{Transport of magnetic field on the sun}.
\bjtitle{Astrophys. J.}
\bvolume{140},
\bfpage{1547}
\doiurl{10.1086/148058}.
\end{barticle}
\endbibitem




\bibitem[\protect\citeauthoryear{Meir \textit{et~al.}}{2004}]{Meir2004}
\begin{barticle}
\bauthor{\bsnm{Meir}, \binits{A.J.}},
\bauthor{\bsnm{Schmidt}, \binits{P.G.}},
\bauthor{\bsnm{Bakhtiyarov}, \binits{S.I.}},
\bauthor{\bsnm{Overfelt}, \binits{R.A.}}:
\byear{2004},
\batitle{Numerical simulation of steady liquid -- metal flow in the presence of
  a static magnetic field}.
\bjtitle{J. Appl. Mech.}
\bvolume{71},
\bfpage{786}
\doiurl{10.1115/1.1796450}.
\end{barticle}
\endbibitem



\bibitem[\protect\citeauthoryear{Okhlopkov}{2014}]{Okhlopkov2014}
\begin{barticle}
\bauthor{\bsnm{Okhlopkov}, \binits{V.P.}}:
\byear{2014},
\batitle{The 11-year cycle of solar activity and configurations of the
  planets}.
\bjtitle{Mosc. U. Phys. B.}
\bvolume{69},
\bfpage{257}
\doiurl{10.3103/S0027134914030126}.
\end{barticle}
\endbibitem



\bibitem[\protect\citeauthoryear{Palus \textit{et~al.}}{2000}]{Palus2000}
\begin{barticle}
\bauthor{\bsnm{Palus}, \binits{M.}},
\bauthor{\bsnm{Kurths}, \binits{J.}},
\bauthor{\bsnm{Schwarz}, \binits{U.}},
\bauthor{\bsnm{Novotna}, \binits{D.}},
\bauthor{\bsnm{Charvatova}, \binits{I.}}:
\byear{2000},
\batitle{Is the solar activity cycle synchronized with the solar inertial
  motion?}
\bjtitle{Int. J. Bifurc. Chaos}
\bvolume{10},
\bfpage{2519}
\doiurl{10.1142/S0218127400001766}.
\end{barticle}
\endbibitem

\bibitem[\protect\citeauthoryear{Parker}{1955}]{Parker1955}
\begin{barticle}
\bauthor{\bsnm{Parker}, \binits{E.N.}}:
\byear{1955},
\batitle{Hydromagnetic dynamo models}.
\bjtitle{Astrophys. J.}
\bvolume{122},
\bfpage{293}
\doiurl{10.1086/146087}.
\end{barticle}
\endbibitem



\bibitem[\protect\citeauthoryear{Pikovsky, Rosenblum, and
  Kurths}{2001}]{Pikovsky}
\begin{bbook}
\bauthor{\bsnm{Pikovsky}, \binits{A.}},
\bauthor{\bsnm{Rosenblum}, \binits{M.}},
\bauthor{\bsnm{Kurths}, \binits{J.}}:
\byear{2001},
\bbtitle{Synchronizations: A universal concept in nonlinear sciences},
\bpublisher{Cambridge University Press}, \blocation{Cambridge}.
\end{bbook}
\endbibitem

\bibitem[\protect\citeauthoryear{Pitts and Tayler}{1985}]{Pittstayler1985}
\begin{barticle}
\bauthor{\bsnm{Pitts}, \binits{E.}},
\bauthor{\bsnm{Tayler}, \binits{R.J.}}:
\byear{1985},
\batitle{The adiabatic stability of stars containing magnetic-fields. 6. The
  influence of rotation}.
\bjtitle{Mon. Not. Roy. Astron. Soc.}
\bvolume{216},
\bfpage{139}
\doiurl{10.1093/mnras/216.2.139}.
\end{barticle}
\endbibitem



\bibitem[\protect\citeauthoryear{Proctor}{2006}]{Proctor2006}
\begin{barticle}
\bauthor{\bsnm{Proctor}, \binits{M.}}:
\byear{2006},
\batitle{Dynamo action and the sun}.
\bjtitle{EAS Publications Series}
\bvolume{21},
\bfpage{241}
\doiurl{10.1051/eas:2006116}.
\end{barticle}
\endbibitem



\bibitem[\protect\citeauthoryear{R\"adler and Stepanov}{2006}]{Raedler2006}
\begin{barticle}
\bauthor{\bsnm{R\"adler}, \binits{K.}},
\bauthor{\bsnm{Stepanov}, \binits{R.}}:
\byear{2006},
\batitle{Mean electromotive force due to turbulence of a conducting fluid in
  the presence of mean flow}.
\bjtitle{Phys. Rev. E}
\bvolume{73},
\bfpage{056311}
\doiurl{10.1103/PhysRevE.73.056311}.
\end{barticle}
\endbibitem



\bibitem[\protect\citeauthoryear{R\"{u}diger, Kitchatinov, and
  Hollerbach}{2013}]{Ruediger2013}
\begin{bbook}
\bauthor{\bsnm{R\"{u}diger}, \binits{G.}},
\bauthor{\bsnm{Kitchatinov}, \binits{L.L.}},
\bauthor{\bsnm{Hollerbach}, \binits{R.}}:
\byear{2013},
\bbtitle{Magnetic processes in astrophysics},
\bpublisher{Wiley-VCH}, \blocation{Berlin}.
\end{bbook}
\endbibitem

\bibitem[\protect\citeauthoryear{R{\"u}diger
  \textit{et~al.}}{2015}]{Ruediger2015}
\begin{botherref}
\oauthor{\bsnm{R{\"u}diger}, \binits{G.}},
\oauthor{\bsnm{Schultz}, \binits{M.}},
\oauthor{\bsnm{Gellert}, \binits{M.}},
\oauthor{\bsnm{Stefani}, \binits{F.}}:
\byear{2015},
\batitle{Subcritical excitation of the current-driven Tayler instability by
  super-rotation}.
\bjtitle{Phys. Fluids}
\bvolume{28},
\bfpage{014105}
\doiurl{10.1063/1.4939270}.
\end{botherref}
\endbibitem




\bibitem[\protect\citeauthoryear{Scafetta}{2010}]{Scafetta2010}
\begin{barticle}
\bauthor{\bsnm{Scafetta}, \binits{N.}}:
\byear{2010},
\batitle{Empirical evidence for a celestial origin of the climate oscillations
  and its implications}.
\bjtitle{J. Atmos. Sol.-Terr. Phys.}
\bvolume{72},
\bfpage{951}
\doiurl{10.1016/j.jastp.2010.04.015}.
\end{barticle}
\endbibitem




\bibitem[\protect\citeauthoryear{Scafetta}{2014}]{Scafetta2014}
\begin{barticle}
\bauthor{\bsnm{Scafetta}, \binits{N.}}:
\byear{2014},
\batitle{The complex planetary synchronization structure of the solar system}.
\bjtitle{Pattern Recogn. Phys.}
\bvolume{2},
\bfpage{1}
\doiurl{10.5194/prp-2-1-2014}.
\end{barticle}
\endbibitem


\bibitem[\protect\citeauthoryear{Schmitt, Sch\"ussler, and Ferriz~Mas}{1996}]{Schmitt1996}
\begin{barticle}
\bauthor{\bsnm{Schmitt}, \binits{D.}},
\bauthor{\bsnm{Sch\"ussler}, \binits{M.}},
\bauthor{\bsnm{Ferriz~Mas}, \binits{A.}}:
\byear{1996},
\batitle{Intermittent solar activity by an on-off dynamo}.
\bjtitle{Astron. Astrophys.}
\bvolume{311},
\bfpage{L1}.
\end{barticle}
\endbibitem



\bibitem[\protect\citeauthoryear{Seilmayer
  \textit{et~al.}}{2012}]{Seilmayer2012}
\begin{barticle}
\bauthor{\bsnm{Seilmayer}, \binits{M.}},
\bauthor{\bsnm{Stefani}, \binits{F.}},
\bauthor{\bsnm{Gundrum}, \binits{T.}},
\bauthor{\bsnm{Weier}, \binits{T.}},
\bauthor{\bsnm{Gerbeth}, \binits{G.}},
\bauthor{\bsnm{Gellert}, \binits{M.}},
\bauthor{\bsnm{R{\"u}diger}, \binits{G.}}:
\byear{2012},
\batitle{Experimental evidence for Tayler instability in a liquid metal
  column}.
\bjtitle{Phys. Rev. Lett.}
\bvolume{108},
\bfpage{244501}
\doiurl{10.1103/PhysRevLett.108.244501}.
\end{barticle}
\endbibitem



\bibitem[\protect\citeauthoryear{Spruit}{2002}]{Spruit2002}
\begin{barticle}
\bauthor{\bsnm{Spruit}, \binits{H.}}:
\byear{2002},
\batitle{Dynamo action by differential rotation in a stably stratified stellar
  interior}.
\bjtitle{Astron. Astrophys.}
\bvolume{381},
\bfpage{923}
\doiurl{10.1051/0004-6361:20011465}.
\end{barticle}
\endbibitem




\bibitem[\protect\citeauthoryear{Steenbeck and Krause}{1969}]{Steenbeck1969}
\begin{barticle}
\bauthor{\bsnm{Steenbeck}, \binits{M.}},
\bauthor{\bsnm{Krause}, \binits{F.}}:
\byear{1969},
\batitle{Zur Dynamotheorie stellarer und planetarer Magnetfelder. I. 
Berechnung sonnen\"ahnlicher Wechselfeldgeneratoren}.
\bjtitle{Astron. Nachr.}
\bvolume{291},
\bfpage{49}
\doiurl{10.1002/asna.19692910201}.
\end{barticle}
\endbibitem



\bibitem[\protect\citeauthoryear{Steenbeck, Krause, and
  R\"adler}{1966}]{Steenbeck1966}
\begin{barticle}
\bauthor{\bsnm{Steenbeck}, \binits{M.}},
\bauthor{\bsnm{Krause}, \binits{F.}},
\bauthor{\bsnm{R\"adler}, \binits{K.-H.}}:
\byear{1966},
\batitle{Berechnung der mittleren {L}orentz-{F}eldst{\"a}rke {v}x{B} f{\"u}r
  ein elektrisch leitendes {M}edium in turbulenter durch
  {C}oriolis-{K}r{\"a}fte beeinflusster {B}ewegung}.
\bjtitle{Z. Naturforsch. A}
\bvolume{21}(\bissue{4}),
\bfpage{369}
\doiurl{10.1515/zna-1966-0401}.
\end{barticle}
\endbibitem


\bibitem[\protect\citeauthoryear{Stefani and Kirillov}{2015}]{Stefani2015}
\begin{barticle}
\bauthor{\bsnm{Stefani}, \binits{F.}},
\bauthor{\bsnm{Kirillov}, \binits{O.N.}}:
\byear{2015},
\batitle{Destabilization of rotating flows with positive shear by azimuthal magnetic fields}.
\bjtitle{Phys. Rev. E}
\bvolume{92},
\bfpage{051001(R)}
\doiurl{10.1103/PhysRevE.92.051001}.
\end{barticle}
\endbibitem




\bibitem[\protect\citeauthoryear{Stix}{1972}]{Stix1972}
\begin{barticle}
\bauthor{\bsnm{Stix}, \binits{M.}}:
\byear{1972},
\batitle{Nonlinear dynamo waves}.
\bjtitle{Astron. Astrophys.}
\bvolume{20},
\bfpage{9}.
\end{barticle}
\endbibitem

\bibitem[\protect\citeauthoryear{Svensmark and
  Friis-Christensen}{1997}]{Svensmark1997}
\begin{barticle}
\bauthor{\bsnm{Svensmark}, \binits{H.}},
\bauthor{\bsnm{Friis-Christensen}, \binits{E.}}:
\byear{1997},
\batitle{Variation of cosmic ray flux and global cloud coverage - a missing
  link in solar-climate relationships}.
\bjtitle{J. Atmos. Sol.-Terr. Phys.}
\bvolume{59},
\bfpage{1225}
\doiurl{10.1016/S1364-6826(97)00001-1}.
\end{barticle}
\endbibitem



\bibitem[\protect\citeauthoryear{Takahashi}{1968}]{Takahashi1968}
\begin{barticle}
\bauthor{\bsnm{Takahashi}, \binits{K.}}:
\byear{1968},
\batitle{On the relation between the solar activity cycle and the solar tidal
  force induced by the planets}.
\bjtitle{Solar Phys.}
\bvolume{3},
\bfpage{598}
\doiurl{10.1007/s11207-005-4086-7}.
\end{barticle}
\endbibitem

\bibitem[\protect\citeauthoryear{Tayler}{1973}]{Tayler1973}
\begin{barticle}
\bauthor{\bsnm{Tayler}, \binits{R.J.}}:
\byear{1973},
\batitle{The adiabatic stability of stars containing magnetic fields-I: Toroidal fields}.
\bjtitle{Mon. Not. Roy. Astron. Soc.}
\bvolume{161},
\bfpage{365}
\doiurl{10.1093/mnras/161.4.365}.
\end{barticle}
\endbibitem


\bibitem[\protect\citeauthoryear{Vainshtein and Cattaneo}{1992}]{Vainshtein1992}
\begin{barticle}
\bauthor{\bsnm{Vainshtein}, \binits{S.I.}},
\bauthor{\bsnm{Cattaneo}, \binits{F.}}:
\byear{1992},
\batitle{Nonlinear restrictions on dynamo action}.
\bjtitle{Astrophys. J.}
\bvolume{393},
\bfpage{165}
\doiurl{10.1086/171494}.
\end{barticle}
\endbibitem


\bibitem[\protect\citeauthoryear{Weber, Fan, and Miesch}{2013}]{Weberfan2013}
\begin{barticle}
\bauthor{\bsnm{Weber}, \binits{M.A.}},
\bauthor{\bsnm{Fan}, \binits{Y.}},
\bauthor{\bsnm{Miesch}, \binits{M.S.}}:
\byear{2013},
\batitle{Comparing simulations of rising flux tubes through the solar
  convection zone with observations of solar active regions: Constraining the
  dynamo field strength}.
\bjtitle{Solar Phys.}
\bvolume{287},
\bfpage{239}
\doiurl{10.1007/s11207-012-0093-7}.
\end{barticle}
\endbibitem



\bibitem[\protect\citeauthoryear{Weber \textit{et~al.}}{2013}]{Weber2013}
\begin{barticle}
\bauthor{\bsnm{Weber}, \binits{N.}},
\bauthor{\bsnm{Galindo}, \binits{V.}},
\bauthor{\bsnm{Stefani}, \binits{F.}},
\bauthor{\bsnm{Weier}, \binits{T.}},
\bauthor{\bsnm{Wondrak}, \binits{T.}}:
\byear{2013},
\batitle{Numerical simulation of the Tayler instability in liquid metals}.
\bjtitle{New J. Phys.}
\bvolume{15},
\bfpage{043034}
\doiurl{10.1088/1367-2630/15/4/043034}.
\end{barticle}
\endbibitem

\bibitem[\protect\citeauthoryear{Weber \textit{et~al.}}{2015}]{Weber2015}
\begin{barticle}
\bauthor{\bsnm{Weber}, \binits{N.}},
\bauthor{\bsnm{Galindo}, \binits{V.}},
\bauthor{\bsnm{Stefani}, \binits{F.}},
\bauthor{\bsnm{Weier}, \binits{T.}}:
\byear{2015},
\batitle{The Tayler instability at low magnetic Prandtl numbers: between chiral
  symmetry breaking and helicity oscillations}.
\bjtitle{New J. Phys.}
\bvolume{17},
\bfpage{113013}
\doiurl{10.1088/1367-2630/17/11/113013}.
\end{barticle}
\endbibitem

\bibitem[\protect\citeauthoryear{Weiss and Tobias}{2016}]{Weiss2016}
\begin{barticle}
\bauthor{\bsnm{Weiss}, \binits{N.O.}},
\bauthor{\bsnm{Tobias}, \binits{S.M}}:
\byear{2016},
\batitle{Supermodulation of the Sun's magnetic activity: the effect of
symmetry changes}.
\bjtitle{Mon. Not. Roy. Astron. Soc.}
\bvolume{456},
\bfpage{2654}
\doiurl{10.1093/mnras/stv2769}.
\end{barticle}
\endbibitem




\bibitem[\protect\citeauthoryear{Wilmot-Smith
  \textit{et~al.}}{2006}]{Wilmotsmith2006}
\begin{barticle}
\bauthor{\bsnm{Wilmot-Smith}, \binits{A.L.}},
\bauthor{\bsnm{Nandy}, \binits{D.}},
\bauthor{\bsnm{Hornig}, \binits{G.}},
\bauthor{\bsnm{Martens}, \binits{P.C.H.}}:
\byear{2006},
\batitle{A time delay model for solar and stellar dynamos}.
\bjtitle{Astrophys. J.}
\bvolume{652},
\bfpage{696}
\doiurl{10.1088/0004-637X/740/2/89}.
\end{barticle}
\endbibitem

\bibitem[\protect\citeauthoryear{Wilson}{2013}]{Wilson2013}
\begin{barticle}
\bauthor{\bsnm{Wilson}, \binits{I.R.G.}}:
\byear{2013},
\batitle{The Venus-Earth-Jupiter spin-orbit coupling model}.
\bjtitle{Pattern Recogn. Phys.}
\bvolume{1},
\bfpage{147}
\doiurl{10.5194/prp-1-147-2013}.
\end{barticle}
\endbibitem



\bibitem[\protect\citeauthoryear{Wood}{1972}]{Wood1972}
\begin{barticle}
\bauthor{\bsnm{Wood}, \binits{K.}}:
\byear{1972},
\batitle{Sunspots and planets}.
\bjtitle{Nature}
\bvolume{240}(\bissue{5376}),
\bfpage{91}
\doiurl{10.1038/240091a0}.
\end{barticle}
\endbibitem

\bibitem[\protect\citeauthoryear{Yoshimura}{1975}]{Yoshimura1975}
\begin{barticle}
\bauthor{\bsnm{Yoshimura}, \binits{H.}}:
\byear{1975},
\batitle{Solar-cycle dynamo wave propagation}.
\bjtitle{Astrophys. J.}
\bvolume{201},
\bfpage{740}
\doiurl{10.1086/153940}.
\end{barticle}
\endbibitem





\bibitem[\protect\citeauthoryear{Zahn, Brun, and Mathis}{2007}]{Zahn2007}
\begin{barticle}
\bauthor{\bsnm{Zahn}, \binits{J.-P.}},
\bauthor{\bsnm{Brun}, \binits{A.S.}},
\bauthor{\bsnm{Mathis}, \binits{S.}}:
\byear{2007},
\batitle{On magnetic instabilities and dynamo action in stellar radiation
  zones}.
\bjtitle{Astron. Astrophys.}
\bvolume{474},
\bfpage{145}
\doiurl{10.1051/0004-6361:20077653}.
\end{barticle}
\endbibitem


\bibitem[\protect\citeauthoryear{Zhang \textit{et~al.}}{2012}]{Zhang2012}
\begin{barticle}
\bauthor{\bsnm{Zhang}, \binits{H.}},
\bauthor{\bsnm{Moss}, \binits{D.}},
\bauthor{\bsnm{Kleeorin}, \binits{N.}},
\bauthor{\bsnm{Kuzanyan}, \binits{K.}},
\bauthor{\bsnm{Rogachevskii}, \binits{I.}},
\bauthor{\bsnm{Sokoloff}, \binits{D.}},
\bauthor{\bsnm{Gao}, \binits{Y.}},
\bauthor{\bsnm{Xu}, \binits{H.}}:
\byear{2012},
\batitle{Current helicity of active regions as a tracer of large-scale solar
  magnetic helicity}.
\bjtitle{Astrophys. J.}
\bvolume{751}
\doiurl{10.1088/0004-637X/751/1/47}.
\end{barticle}
\endbibitem


\bibitem[\protect\citeauthoryear{Zhang \textit{et~al.}}{2003}]{Zhang2003}
\begin{barticle}
\bauthor{\bsnm{Zhang}, \binits{K.}},
\bauthor{\bsnm{Chan}, \binits{K.H.}},
\bauthor{\bsnm{Zou}, \binits{J.}},
\bauthor{\bsnm{Liao}, \binits{X.}},
\bauthor{\bsnm{Schubert}, \binits{G.}}:
\byear{2003},
\batitle{A three-dimensional spherical nonlinear interface dynamo}.
\bjtitle{Astrophys. J.}
\bvolume{596},
\bfpage{663}
\doiurl{10.1086/377600}.
\end{barticle}
\endbibitem



\end{thebibliography}

\end{article} 

\end{document}